# Content-Priority based Interest Forwarding in Content Centric Networks

Muhammad Aamir, *Member IEEE*
National Bank of Pakistan, Karachi, Pakistan
aamir.nbpit@yahoo.com

*Abstract*—**Content Centric Networking (CCN) is a recent advancement in communication networks where the current research is mainly focusing on routing & cache management strategies of CCN. Nonetheless, other perspectives such as network level security and service quality are also of prime importance; areas which have not been covered deeply so far. This paper introduces an interest forwarding mechanism to process the requests of consumers at a CCN router. Interest packets are forwarded with respect to the priorities of addressed content while the priority level settings are done by content publishers during an initialization phase using a collaborative mechanism of exchanging messages to agree to the priority levels of all content according to the content-nature. Interests with higher priority content are recorded in Pending Interest Table (PIT) as well as forwarded to content publishers prior to those with lower priority content. A simulation study is also conducted to show the effectiveness of proposed scheme and we observe that the interests with higher priority content are satisfied earlier than the interests with lower priority content.**

## I. INTRODUCTION

Content Centric Networking (CCN) is a paradigm shift of traditional IP based communication towards content oriented communication. In this model, the hosts are not addressed using their IP address to retrieve the content. Instead, content is directly addressed using naming conventions representing the hierarchy from top to bottom where the desired content is located. Two packet types exist in CCN i.e. interest packet and data packet [1]. Interest packet is a request (initiated by a *consumer*) that contains the content name to be addressed. Data packet is the response (generated by a *publisher*) of interest packet which satisfies the requirements of the corresponding interest. The content names are aggregated by prefix for scalability and are globally routable [2]. An example may be considered as: */bank/data/account/balance*. Here we assume that a bank's content server provides the balance statement against an account holder's corresponding interest packet. **/bank/data** is the prefix which can be used by different account holders in order to reach the bank's content server. The account and balance query (**/account/balance**) however varies for each account holder and the corresponding values are placed in respective interest packets. The CCN technology, also known to be Named Data Networking (NDN) due to the fact that the content (or data) is *named* through interest packets, is one of the proposals of Information Centric Networking (ICN) [3]. A data packet in CCN satisfies consumer's interest packet if content name in interest packet is a prefix of content name in data packet. The CCN router works with three major processing sections. They are Pending Interest Table (PIT), Content Store (CS) and Forwarding Information Base (FIB). The normal communication flow may be described as: Interest from a consumer is received at an interface and checked in CS whether the content is already cached and available. If the content is available, it is sent to the consumer and interest is said to be consumed. When the content is not available in CS, it is checked in PIT whether the same content has previously been asked through the router but it is unanswered so far. If a corresponding entry is found in PIT with the same content name prefix, the interest arrival interface is recorded in the PIT list. If the corresponding entry is not found in PIT, the request is forwarded to FIB where it checks its forwarding table for the requested prefix. If the content name prefix is matched, the interest is propagated to upstream nodes (routers or the ultimate content publisher) by FIB through connected interfaces while the interest entry is also added in PIT. On receiving data from the upstream node, some conditions are checked whether newly arrived data is not still available in CS and prefix matching entries are still held in FIB and PIT (any mismatch or negative result forces FIB to discard data to avoid duplication and reception of unsolicited data packets). When all checks are performed and arrived data is found to be legitimate for processing to satisfy a pending interest, the content is validated and cached in the CS. The content is then sent to each interest arrival interface[1] found in the PIT list of interfaces against the corresponding interest. In figure 1, the sequences of interest forwarding and content retrieval are shown.

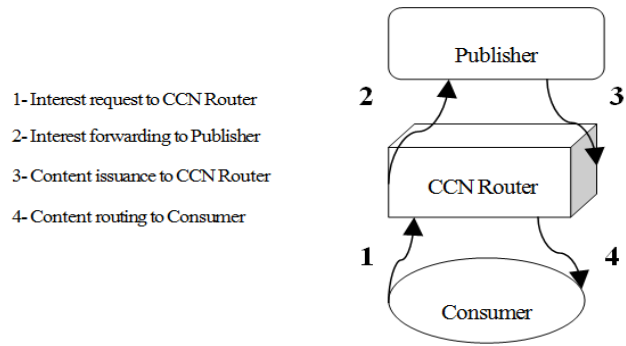

Figure 1. Sequences of interest forwarding and content retrieval.

---

[1] In CCN, the term 'interface' may be replaced by **face** [1]. We also use the term **'face'** in next sections of this paper.

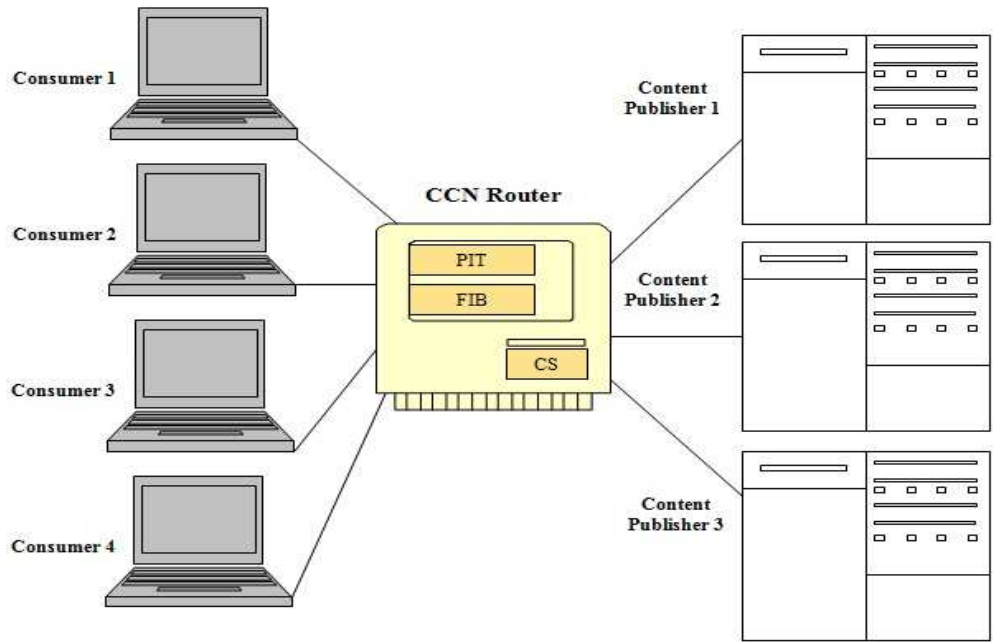

Figure 2.  Prototype network (CCN) of proposed scheme.

## II.  RELATED WORK

The Quality of Service (QoS) is always an important consideration when differentiated services are the part of a network in terms of priorities. The current CCN research is mainly focusing on routing and cache management strategies whereas QoS related area has not been covered deeply so far. Some QoS focused work of CCN discusses intelligent path selection through FIB based on route congestion in networks having more than one path available to reach one or more content sources for the same type of content [4]. Bandwidth sharing is also discussed to manage low latency requirements of delay sensitive flows [5]. The proposal of Greedy Ant Colony Forwarding (GACF) algorithm also exists for intelligent and efficient interest forwarding over multiple routes [6]. Some other related work covers the aspect of efficient resource sharing in terms of cache storage and bandwidth utilization keeping in view some or all the parameters of content availability, content popularity, network topology, cache size and scarcity of network resources [7-10]. However, the aspect of content priority is yet to be explored which is different from the content popularity feature. In this paper, we describe an interest forwarding mechanism to process the content based requests at CCN router where the interest packets are forwarded with respect to the assigned priorities of addressed content. Interests with higher priority content are recorded in PIT on top of those with lower priority content as well as they are forwarded to content publishers on priority. We do not consider networks with multiple routes for reaching one or more content sources for the same type of content.

## III.  CONTENT PRIORITY BASED INTEREST FORWARDING

In order to discuss the proposed approach, we take a prototype network of four consumers, one CCN router and three content publishers as shown in figure 2. We consider an initialization phase in which all content publishers announce their longest matching prefixes as well as the corresponding priority values of the content they contain. For a successful interest forwarding mechanism, we assume that the priority level settings must follow a global rule within the network. The announcements are made from content publishers to the downstream router (*CCN Router* in figure 2). In case of more than one router, a router receiving the announcement assumes the responsibility of forwarding the same to further downstream router(s). The process continues until the last router (one which is directly connected to the consumer node or LAN) receives the announcement. We consider that the mentioned content publishers of figure 2 provide different content types in the quantity of two each. Table I indicates the announced content prefixes with respective content priorities (highest number → highest priority). We further consider interest queues within CCN router for each of the mentioned consumers. The queues are formed in parallel (using an allocated space of available cache memory) and interest packets are queued in the same sequence as they are generated by consumers. At the processing edge of queues, we insert a priority check pointer that takes into consideration all interests (one from each queue) lying at the edge. At this point, a check is performed for interests asking for the content with respect to the announced priorities from content publishers. A consumer asking for the content of the highest priority amongst examined interests is given preference. Hence, its PIT information is recorded and the interest is forwarded to the respective content publisher on priority (this forwarding of interest has to follow CCN routing rules i.e. forwarded through FIB using longest matching prefix provided the prefix is available in FIB routing information). The next interest then takes the place on the edge of the queue from which the prior interest is just forwarded. Similarly, the check is again performed for priorities of content and the next interest is also forwarded for the highest priority content of this cycle. In this way, consumers asking for the prior content are given preferences in PIT information recording and interest forwarding to content publishers. Consequently, they receive data (content) before other consumers who ask for relatively low priority content. In order to avoid any mishandling or malicious node behavior, we do not give authority of setting priority levels of content to ultimate consumers. We therefore consider that the priority level settings are done by content publishers using a collaborative mechanism of exchanging messages to agree to the priority levels of all content according to the content-nature. The agreed levels are finally recorded in CCN router to make decisions on the priority interests. Alternatively, there may be a dedicated authorized mechanism to follow a global rule according to the content-nature within the network. The benefit of this priority marking is that the content of critical or sensitive nature can be delivered to the corresponding consumers earlier than that of ordinary nature in the occurrences of processing cycles during which they are forwarded. For example, it is depicted in table I that the content of disease cases hosted by a hospital server (publisher) will have the priority over other types of content in a processing cycle due to relatively higher priority markings globally within the network. Consequently, it is to be delivered earlier than video, image and other content.

TABLE I. CONTENT PREFIXES AND PRIORITIES IN PROTOTYPE NETWORK

| Content Publisher | Content Prefix | Content Priority |
|---|---|---|
| Content Publisher 1 | /data/hospital/cancer_cases | 6 |
| Content Publisher 1 | /data/hospital/heart_cases | 5 |
| Content Publisher 2 | /data/bank/transactions | 4 |
| Content Publisher 2 | /data/bank/services | 3 |
| Content Publisher 3 | /data/web_server/mp4_videos | 2 |
| Content Publisher 3 | /data/web_server/bmp_images | 1 |

In figure 3, it is shown that the *Priority Check Pointer* checks all interests (one from each queue) lying at the processing edge of *CCN Router*. We consider a case where interest packets are present at the processing edge of all four interest queues shown in figure 3. The **Consumer 1** needs */data/bank/transactions* (Priority Value: 4). Similarly, **Consumer 2** needs */data/web_server/dat_videos* (Priority Value: 2), **Consumer 3** needs */data/hospital/cancer_cases* (Priority Value: 6) and **Consumer 4** needs */data/bank/services* (Priority Value: 3). In such a case, using table I, our priority check pointer will check the asked content's priorities and process the request of *Consumer 3* as the relevant content priority is the highest in this cycle. Similarly, next candidate can be *Consumer 1* if the request from *Consumer 3* demands a content of priority lower than the priority of content required by *Consumer 1* in the next cycle. It implies that the priority consideration is made only at the processing edge of CCN router for one interest per consumer in a processing cycle. Therefore, the sequence of interest generation is an important factor for a consumer as the priority interests generated later by a consumer may be lying far from the processing edge of CCN router after some low priority interests of the same consumer. It is understood that an interest is immediately satisfied if the desired content is already available in CS of the router. An interest of this kind is never forwarded to FIB anymore unless the relevant content is removed from CS. Hence, the proposed scheme is not required for such interest packets as they do not occupy the processing queues in our model.

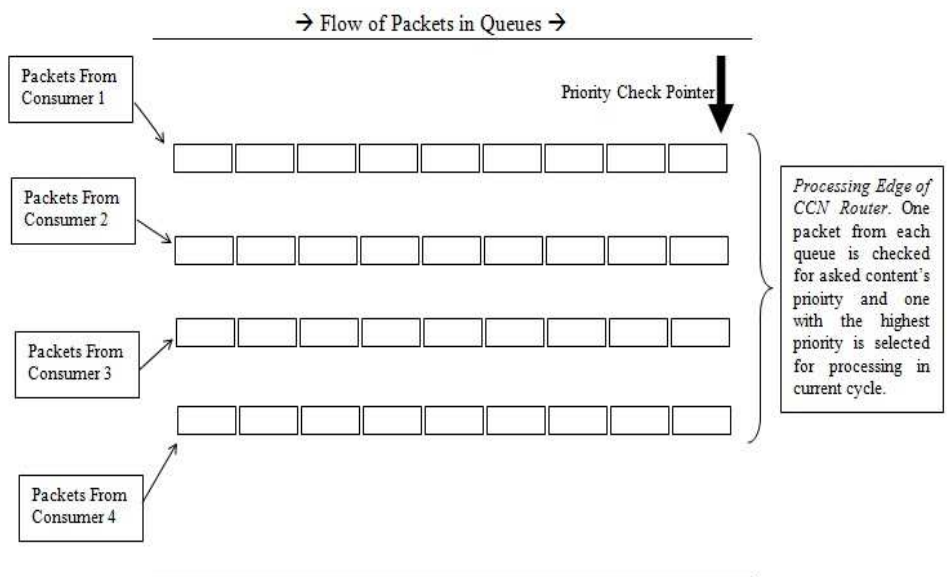

Figure 3. Packet (Interest) queues in CCN router for priority check.

We now consider some other scenarios in our proposed scheme under which the described model can also provide the desired results. For example, we may have interest packets in only one queue due to the reason that a single consumer asks for some content whereas the others are silent. The packets are hence sequentially processed as they arrive on the processing edge. Further, we may also have interest packets in some queues due to the reason that a few consumers ask for content whereas the others are silent. The scheme is therefore applied on the interest containing queues whereas empty queues are ignored. Another situation may arise when two or more interests at the processing edge can ask for the same type of content having the highest priority as compared to the content asked by other interests in the same cycle. Here the priority check pointer forwards one interest at random through FIB while other interests in the same group (demanding the same highest priority content) are dropped from respective queues. However, all consumers who ask for the content are added in PIT. It ensures that all consumers who need the content get data when provided by the content publisher as all corresponding faces are listed in PIT. It is important that two different types of content cannot have the same priority level in a network. In order to ensure that no queue is ignored for a longer period of time due to low priority content requirements while other queues demand relatively higher priority content, we consider that an interest from a queue must not remain stuck at the processing edge after 25 interests are processed from other queues. Hence we implement a counter with each queue which is incremented by 1 when an interest is forwarded from another queue. The counter is reset to 0 when an interest is forwarded from the same queue on priority basis. If the counter reaches the value of 25 at a point of time, the priority check pointer stops checking priorities for the current cycle and the interest is forwarded from the queue having counter value of 25. The counter of the queue is then reset to 0 and the priority check pointer again starts to check priorities from the next cycle. If two or more queues reach counter value of 25 one after the other, the priority check pointer thus stops checking priorities for 'n' number of cycles (where 'n' is the number of queues reaching counter value of 25) and the interests are forwarded from such queues one by one while resetting counters to 0 accordingly. For the reliability purpose, we avoid assuming any situation where some of the mentioned consumers can act as misbehaving agents by repeatedly asking for higher priority content in sequential manner to generate a denial of service effect for legitimate consumers [11-12]. This situation may be worked upon in future extensions of our work. Nonetheless, the CCN's characteristic of *caching the content* provides a level of inbuilt security against this threat.

## IV. PERFORMANCE EVALUATION

Using the prototype network of figure 2 and priorities of content mentioned in table I, we randomly generate 100 interests using 4 consumers in 10 seconds of simulation. The scenario is built in NS2 simulator using consumer, router and publisher objects developed in C++. We first obtain the patterns of interest generation and content retrieval during the simulation time. We record interest generation as soon as an interest is forwarded from the processing edge of CCN router through FIB. The content retrieval is recorded when the first content chunk is received at the corresponding face of CCN router. The statistics are shown in figure 4 and we observe that the content of respective priorities is retrieved within a reasonable time once the corresponding interests are forwarded by CCN router. However, some delays are definitely involved depending on the interest creation time, nature of the required content and the time taken by the packets to remain in queues within CCN router. Some initial delay also involves the initialization phase in which all content publishers announce their longest matching prefixes as well as the corresponding priority values of the content they contain. The priority level settings are done by content publishers using a collaborative mechanism of exchanging messages to agree to the priority levels of all content according to the content-nature (mentioned in table I). The agreed levels are finally recorded in CCN router to make decisions on the priority interests. Therefore, it is relevant to obtain delay statistics to find the effectiveness of our proposed scheme in terms of content priorities.

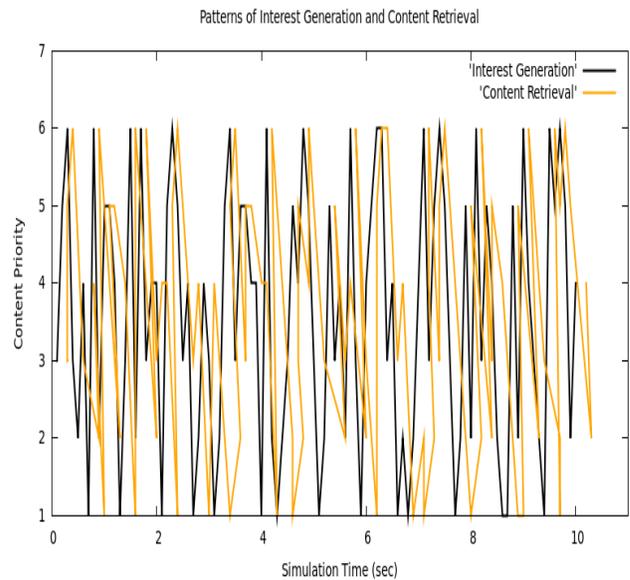

Figure 4. Patterns of Interest Generation and Content Retrieval.

## A. Queuing Delay

In figure 5, the average queuing delay of interest packets in CCN router is shown. The simulation is run for 20 times to minimize simulation error if any and the result is obtained on average. It is observed that the average queuing delay of interest packets in CCN router is decreased with the increasing content priorities. It conveys that the interest packets asking for the priority content experience less queuing delays and they are forwarded to the corresponding publishers on priority. It shows the effectiveness of the proposed scheme. As the priority consideration is made only at the processing edge of CCN router for one interest per consumer in a processing cycle, we start calculating the queuing delays of packets once they arrive at the processing edge of CCN router.

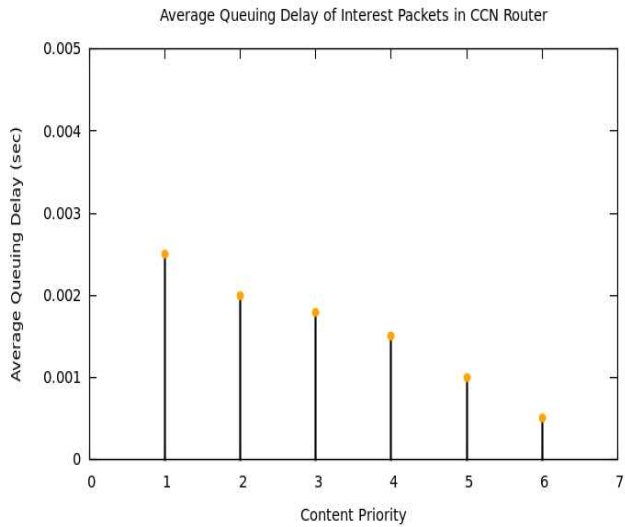

Figure 5. Average Queuing Delay of Interest Packets in CCN Router.

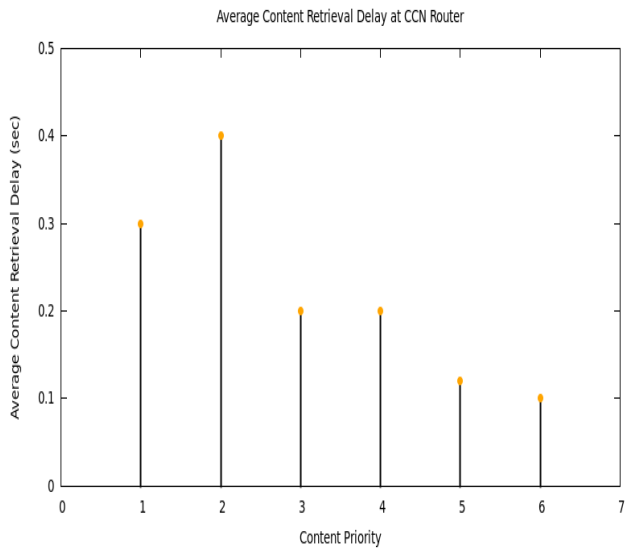

Figure 6. Average Content Retrieval Delay at CCN Router.

## B. Content Retrieval Delay

In the next simulation run, we obtain the average content retrieval delay at CCN router. It is the delay involved in retrieving the content at the corresponding face of CCN router after the respective interest is forwarded to the publisher. The content retrieval is recorded when the first content chunk is received at the corresponding face of the router. In figure 6, it is shown that the average content retrieval delay is generally decreased with the increasing content priorities. It conveys that the priority content is retrieved at CCN router earlier than the ordinary content as the corresponding interest packets also experience less queuing delays and forwarded to the publishers on priority. It shows the effectiveness of the proposed scheme. However, we observe that the content of priority level 2 experiences a higher content retrieval delay than the content of priority level 1. It depicts that the nature and magnitude of content also play their roles accordingly. In our simulation, the content of priority level 1 is composed of bitmap images where a file (content) may have the file size of a few megabytes. On the other hand, the content of priority level 2 is based on mp4 video files where the file sizes have a range from a few megabytes to hundreds of megabytes. Therefore, the larger content takes a considerable time to respond to an interest while providing a comparatively larger content chunk to the router. As the other content having higher priorities in the simulated scenario are usually text files with file sizes of a few kilobytes, they exhibit a general trend i.e. high priority content is retrieved earlier than the low priority content.

## C. Interest Satisfaction Ratio

For the simulated scenario, we also obtain the interest satisfaction ratio i.e. the percentage of interests satisfied during the simulation time of 10 seconds. It is observed that 97% of interest packets are satisfied on an average scale while the simulation is run for 20 times to minimize simulation error if any. The dropped content chunks are those of image and video content as large content may occasionally drop due to high bandwidth requirements. They may also be dropped at router's face if the corresponding PIT entry is removed on reaching its expiry (every PIT entry has its timeout limit) before the content's arrival at the router. The arrival of a larger content takes a considerable time to respond to an interest. Hence, a PIT entry may expire before the corresponding content's arrival at the router. However, it shows that the proposed scheme is effective enough to successfully achieve the desired results by exhibiting an excellent interest satisfaction ratio and experiencing lesser queuing and content retrieval delays for the high priority content.

## V. CONCLUDING REMARKS & FUTURE WORK

In this paper, we introduce an interest forwarding mechanism to process the requests of consumers at a CCN router. Interest packets are forwarded with respect to the priorities of addressed content. The priority level settings are done by content publishers during an initialization phase using a collaborative mechanism of exchanging messages to agree to the priority levels of all content according to the content-nature. We further simulate our prototype network and observe that the interests with higher priority content are recorded in Pending

Interest Table (PIT) as well as forwarded to content publishers prior to those with lower priority content. Hence, the interests with higher priority content are satisfied earlier than the interests with lower priority content. The simulation results show that the proposed scheme is effective enough to successfully achieve the desired results by exhibiting an excellent interest satisfaction ratio and experiencing lesser queuing and content retrieval delays for the high priority content. In future work, we plan to simulate more scenarios by varying the parameters of number of consumers, network size in terms of routers and publishers, nature of the content and interest packets in some queues while keeping the other queues empty in order to obtain various simulation results under different operating conditions. We also plan to derive mathematical relations involved in the mechanism of exchanging messages among content publishers during the initialization phase as well as the relations in the interest forwarding and content retrieval phases.

REFERENCES

[1] V. Jacobson, D.K. Smetters, J.D. Thornton, M. Plass, N. Briggs, and R. Braynard, "Networking Named Content," *Communications of the ACM*, vol. 55, no. 1, pp. 117-124, January 2012.

[2] J.S. Khoury and C.T. Abdallah, "A Survey of Novel Internetwork (and Naming) Architectures," *Internet Naming and Discovery, Signals and Communication Technology (Chapter 2)*, Springer-Verlag London, 2013, pp. 13-33.

[3] B. Ahlgren, C. Dannewitz, C. Imbrenda, D. Kutscher, and B. Ohlman, "A Survey of Information-Centric Networking," *IEEE Communications Magazine*, vol. 50, no. 7, pp. 26-36, July 2012.

[4] A.Z. Khan, S. Baqai, and F.R. Dogar, "QoS Aware Path Selection in Content Centric Networks," *Proc. of IEEE Int'l Conf. on Communications (ICC)*, pp. 2645-2649, June 2012.

[5] S. Oueslati, J. Roberts, and N. Sbihi, "Flow-aware Traffic Control for a Content-centric Network," *Proc. of IEEE INFOCOM*, pp. 2417-2425, March 2012.

[6] C. Li, W. Liu, and K. Okamura, "A Greedy Ant Colony Forwarding Algorithm for Named Data Networking," *Proc. of APAN*, 2012, pp.17-26.

[7] G. Carofiglio, M. Gallo, L. Muscariello, and D. Perino, "Modeling Data Transfer in Content-Centric Networking," *Proc. of 23$^{rd}$ Int'l Teletraffic Congress (ITC)*, ACM, pp. 111-118, September 2011.

[8] M. Tortelli, I. Cianci, L.A. Grieco, G. Boggia, and P. Camarda, "A Fairness Analysis of Content Centric Networks," *Proc. of IEEE Int'l Conf. on the Network of the Future (NOF)*, pp. 117-121, November 2011.

[9] G. Carofiglio, V. Gehlen, and D. Perino, "Experimental Evaluation of Memory Management in Content-Centric Networking," *Proc. of IEEE Int'l Conf. on Communications (ICC)*, pp. 1-6, June 2011.

[10] G. Carofiglio, M. Gallo, and L. Muscariello, "Bandwidth and Storage Sharing Performance in Information Centric Networking," *Proc. of ACM SIGCOMM Workshop on Information-Centric Networking (ICN)*, pp. 26-31, August 2011.

[11] M. Aamir and M.A. Zaidi, "A Survey on DDoS Attack and Defense Strategies: From Traditional Schemes to Current Techniques." *Interdisciplinary Information Sciences*, vol. 19, no. 2, pp. 173-200, November 2013.

[12] M. Aamir and S.M.A. Zaidi, "Denial-of-Service in Content Centric (Named Data) Networking: A Tutorial and State-of-the-Art Survey." *Security and Communication Networks*, accepted publication, October 2014.